\documentclass[twocolumn,preprintnumbers,amsmath,amssymb]{revtex4}
\usepackage{txfonts}
\usepackage{amsmath}
\usepackage{stmaryrd}
\usepackage{amssymb}
\usepackage{multirow}
\usepackage{cases}
\usepackage[none]{hyphenat}

\usepackage{graphicx}
\usepackage{dcolumn}
\usepackage{bm}
\usepackage{braket} 

\def\beq{\begin{equation}}
\def\eeq{\end{equation}}
\def\beqa{\begin{eqnarray}}
\def\eeqa{\end{eqnarray}}
\newcommand{\eq}[1]{Eq.(#1)}

\begin{document}
\title{Efficient excitation of a two level atom by a single photon in a propagating mode}
\author{Yimin~Wang$^1$, Ji\v{r}\'{i}~Min\'{a}\v{r}$^1$, Lana~Sheridan$^1$ and Valerio~Scarani$^{1,2}$}
\affiliation{$^1$Centre for Quantum Technologies, National University of Singapore, Singapore\\
$^2$Department of Physics, National University of Singapore, Singapore}


\begin{abstract}
State mapping between atoms and photons, and photon-photon interactions play an important role in scalable quantum information processing. We consider the interaction of a two-level atom with a quantized \textit{propagating} pulse in free space and study the probability $P_e(t)$ of finding the atom in the excited state at any time $t$. This probability is expected to depend on (i) the quantum state of the pulse field and (ii) the overlap between the pulse and the dipole pattern of the atomic spontaneous emission. We show that the second effect is captured by a single parameter $\Lambda\in[0,8\pi/3]$, obtained by weighting the dipole pattern with the numerical aperture. Then $P_e(t)$ can be obtained by solving time-dependent Heisenberg-Langevin equations. We provide detailed solutions for both single photon Fock state and coherent states and for various temporal shapes of the pulses.
\end{abstract}


\maketitle


\section{Introduction}
Light-matter interface in free space at quantum level lies at the heart of quantum networks and quantum communication as well as being the fundamental question in quantum optics, and may be less technologically demanding than typical cavity quantum electro-dynamical systems. Recently, achievements have been made for atoms \cite{vanEnk_2000,Pinotsi_2008,Zumofen_2008,Tey_2008,Slodicka_2010}, single molecules \cite{Gerhardt_2007,Wrigge_2008} and quantum dots \cite{Vamivakas_2007}.
Especially, high numerical aperture optics \cite{Sondermann_2009} have been recognized as a key element for free space atom-light coupling and precision spectroscopy, involving fixed aspheric lenses \cite{Tey_2008,Slodicka_2010,Tey_2009,Aljunid_2009,Schuck_2010}, deep parabolic mirrors \cite{Lindlein_2007,Sondermann_2007,Stobinska_2009}, spherical mirrors \cite{Hetet_2010} and phase Fresnel lenses \cite{Streed_2011}.

Regarding the atom-light interaction, there are two different phenomena: the scattering of the light by the atom \cite{Zumofen_2008,Tey_2008,Wrigge_2008} and the absorption of the light by the atom \cite{Pinotsi_2008,Stobinska_2009,Heugel_2010}. Efficient information transfer between atoms and photons requires controlled photon absorption with high probability. However, the time reversal argument implies the main properties of the excitation pulse. These are (i) the spatial profile of the pulse should match the atomic dipole emission pattern, and (ii) the temporal shape of the photon has to be a time inverted version of the spontaneously emitted photon. This means that the atom must be illuminated from all directions by a photon with a rising exponential temporal envelope \cite{Stobinska_2009,Heugel_2010}. Strong focusing can give rise to increased overlap of the light beam with the atomic dipole and this improvement of the atom-light coupling in free space has been predicted by theory \cite{vanEnk_2000,Tey_2009} and demonstrated in experiment \cite{Tey_2008,Slodicka_2010,Aljunid_2009}.

In this paper, we focus on the effect of the temporal-spectral features of the pulse on the probability of finding the atom in the excited state starting from the ground state (``excitation probability"). We present the general formalism and apply it into two specific situations: single photon wave packets and coherent state wave packets.

The paper is organized as follows. In Sec. \ref{s2}, we review a general quantized model of the interaction between an atom and a continuum propagating pulse in free space and introduce the basic parameters describing the coupling between the atomic dipole and the pulse. In Sec. \ref{s3}, we introduce a special case. In Sec. \ref{s4}, the dependence of the atomic excitation probability on the temporal and spectral features of single photon wave packets and coherent state wave packets are investigated, respectively. The excitation probability for realistic focusing setups is discussed in Sec.\ref{s5}. And our results are briefly summarized in Sec.\ref{s6}.


\section{General model and approach}
\label{s2}
We start by considering the interaction of a two-level atom sitting at position $\textbf{r}_a$ with the quantized radiation field with continuum modes in free space.
In Coulomb gauge, the positive-frequency parts of the electric field operators can be expanded as \cite{Smith_2007},
\begin{eqnarray}
\hat{\textbf{E}}^{(+)}(\textbf{r},t)=i \sum_\lambda \int d^3 \textbf{k} \sqrt{\frac{\hbar \omega_k}{{(2 \pi)}^3 2 \epsilon_0}} \hat{a}_{\textbf{k},\lambda}\bm{\epsilon}_{\textbf{k},\lambda} u_{\textbf{k},\lambda} (\textbf{r}) e^{-i\, \omega_k t},
\label{elec1}
\end{eqnarray}
where $\omega_k=c|\textbf{k}|$, $c$ is the vacuum speed of light, $\epsilon_0$ is the permittivity of the vacuum, $\bm{\epsilon}_{\textbf{k},\lambda}(\lambda=1,2)$ are unit polarization vectors, $\bm{\epsilon}_{\textbf{k},\lambda}\cdot\bm{\epsilon}_{\textbf{k},\lambda'}=\delta_{\lambda\lambda'}$, $\bm{\epsilon}_{\textbf{k},\lambda}\cdot \textbf{k}=0$, and the field operators follow the usual commutation relation
\begin{eqnarray}
	\left[ \hat{a}_{\textbf{k},\lambda}, \hat{a}^\dagger_{\textbf{k}',\lambda'} \right] = \delta(\textbf{k} - \textbf{k}')\,\delta_{\lambda,\lambda'}.
\end{eqnarray}

Energy conservation implies the normalization of the spatial mode functions $u_{\textbf{k},\lambda} (\textbf{r})$
\begin{eqnarray}
  \int d^3 \textbf{r} \,\,u^*_{\textbf{k},\lambda} (\textbf{r})\cdot u_{\textbf{k}',\lambda'} (\textbf{r})=\delta(\textbf{k}-\textbf{k}')\,\delta_{\lambda\lambda'}.
\end{eqnarray}

In the interaction picture and rotating-wave approximation, the dynamics of the system is described by the Hamiltonian
\begin{eqnarray}
	\hat{\textbf{H}}_I = -i \hbar \sum_\lambda \int d^3 \textbf{k} \left[g_{\textbf{k},\lambda}(\textbf{r}_a)\,\hat{\sigma}_+ \hat{a}_{\textbf{k},\lambda} e^{-i\,(\omega_k-\omega_a)t}-h.c.\right],
\end{eqnarray}

\noindent where $\omega_a=E_e-E_g$ is the atomic transition frequency and $\hat{\sigma}_+=|e\rangle\langle g|,  \hat{\sigma}_-=|g\rangle\langle e|,  \hat{\sigma}_z=|e\rangle\langle e|-|g\rangle\langle g|=\hat{\sigma}_{+}\hat{\sigma}_{-}-\hat{\sigma}_{-}\hat{\sigma}_{+}$ are atomic operators.
The coupling strength is given by
\begin{eqnarray}
	g_{\textbf{k},\lambda}(\textbf{r}_a)=d\,\sqrt{\frac{\omega_k}{{(2 \pi)}^3 2  \hbar \epsilon_0}}\, u_{\textbf{k},\lambda} (\textbf{r}_a)\,\textbf{e}_d \cdot \bm{\epsilon}_{\bf{k},\lambda},
	\label{eq coupling}
\end{eqnarray}
where $d$ is the value of the dipole momentum and $\textbf{e}_d$ is the unit dipole vector.

The evolution of the operators is governed by a set of coupled Heisenberg equations
\begin{eqnarray}
	\dot{\hat{a}}_{\textbf{k},\lambda}=g^*_{\textbf{k},\lambda}(\textbf{r}_a) \hat{\sigma}_{-} e^{i\,(\omega_k-\omega_a)t} \label{dd eq1},
\end{eqnarray}
	\vskip -0.9cm
	\begin{multline}
	\label{dd eq2}
	\dot{\hat{\sigma}}_{z}=-\Gamma'(\hat{\sigma}_{z}+1)\\
	-2 \sum_\lambda \int d^3 \textbf{k} \left[ g_{\textbf{k},\lambda}(\textbf{r}_a)\hat{\sigma}_{+} \hat{a}_{\textbf{k},\lambda} e^{-i\,(\omega_k-\omega_a)t}+h.c.\right]+\hat{\zeta}_z,
	\end{multline}
	\vskip -0.5cm
\begin{eqnarray}
	\dot{\hat{\sigma}}_{-}= -\frac{\Gamma'}{2}\hat{\sigma}_{-}+\hat{\sigma}_{z}\sum_\lambda \int d^3 \textbf{k} g_{\textbf{k},\lambda}(\textbf{r}_a) 									 \hat{a}_{\textbf{k},\lambda}e^{-i\,(\omega_k-\omega_a)t}+\hat{\zeta}_-, \label{dd eq3}
\end{eqnarray}

\noindent in which the decay term proportional to $\Gamma'$ and the noise operators $\hat{\zeta}$ are introduced to account for the interaction of the atom with the environment.

By integrating Eq.(\ref{dd eq1}), the field operator is decomposed into a free field part and a part radiated by the atom \cite [p. 393]{Tannoudji_2004}
\begin{eqnarray}	
\hat{a}_{\textbf{k},\lambda}(t)=\hat{a}_{\textbf{k},\lambda}(t_0)+g^*_{\textbf{k},\lambda}(\textbf{r}_a)\int_{t_0}^t {\hat{\sigma}_{-}(t')e^{i\,(\omega_k-\omega_a)t'}dt'}.
	\label{a2}
\end{eqnarray}
The substitution of Eq.(\ref{a2}) back into Eq.(\ref{dd eq2}) and (\ref{dd eq3}) gives a set of modified optical Bloch equations \cite{Domokos_2002},
\vskip -0.5cm
\begin{multline}
	\label{z1}
	\dot{\hat{\sigma}}_{z}=-\Gamma(\hat{\sigma}_{z}+1)\\
	-2 \sum_\lambda \int d^3 \textbf{k} \left[ g_{\textbf{k},\lambda}(\textbf{r}_a) \hat{\sigma}_{+} \hat{a}_{\textbf{k},\lambda}(t_0) e^{-i\,(\omega_k-\omega_a)t} + h.c.\right]+\hat{\zeta}_z,
\end{multline}
\vskip -0.7cm
\begin{multline}
	\dot{\hat{\sigma}}_{-}= -\frac{\Gamma}{2}\hat{\sigma}_{-} \\
	+ \hat{\sigma}_{z}\sum_\lambda \int d^3 \textbf{k} g_{\textbf{k},\lambda}(\textbf{r}_a) \hat{a}_{\textbf{k},\lambda}(t_0)e^{-i\,(\omega_k-\omega_a)t}+\hat{\zeta}_-,
	\label{c1}
\end{multline}

\noindent where the standard spontaneous decay rate in free space is made up of two parts \cite{Silberfarb_2004}: $\Gamma=\Gamma'+ \Gamma_{p}$, the decay into the environment $\Gamma'$, which is the non-pulse mode in our case, and the decay to the pulse mode $\Gamma_p$.

Following the same reasoning for the environment field operators $\hat{b}_{\textbf{k}',\lambda'}$ as for the field operators $\hat{a}_{\textbf{k},\lambda}$, one readily finds the explicit form of the noise operators
	\begin{eqnarray}
\hat{\zeta}_z &=& -2 \sum_{\lambda'} \int d^3 \textbf{k}' \left[ g_{\textbf{k}',\lambda'}(\textbf{r}_a)\hat{\sigma}_{+} \hat{b}_{\textbf{k}',\lambda'}(t_0) e^{-i\,(\omega_{k'}-\omega_a)t}+h.c.\right], \label{eq zeta_z} \\
	\hat{\zeta}_{-} &=& \hat{\sigma}_{z}\sum_{\lambda'} \int d^3 \textbf{k}' g_{\textbf{k}',\lambda'}(\textbf{r}_a) \hat{b}_{\textbf{k}',\lambda'}(t_0) e^{-i\,(\omega_{k'}-\omega_a)t},
	\label{eq zeta_-}
\end{eqnarray}
where $g_{\textbf{k}',\lambda'}(\textbf{r}_a)$ is the corresponding coupling strength to the atom.

Furthermore, with the use of the Weisskopf-Wigner theory \cite[p. 207]{Scully_1997}, the explicit formula for the $\Gamma_p$ can be found and is given by,
\begin{eqnarray}
	\Gamma_{p}=2\pi \sum_\lambda \int d^3 \textbf{k} \,{|g_{\textbf{k},\lambda}(\textbf{r}_a)|}^2 \delta(\omega_k-\omega_a).
	\label{eq Gamma_p}
\end{eqnarray}
\noindent Substituting for $g_{{\bf k},\lambda}$ from \eq{\ref{eq coupling}} and going to the spherical coordinates, one gets
\beqa
	\Gamma_p &=& \frac{1}{2 (2\pi)^2} \left( \frac{\omega_a}{c} \right)^3 \frac{d^2}{\hbar \epsilon_0} \sum_{\lambda} \int {\rm d}\Omega \left| u_{{\bf k}_a,\lambda} ({\bf r}_a) \right|^2 |{\bf e}_d \cdot {\bf \epsilon_{{\bf k}_a,\lambda}}|^2  \nonumber \\
					 &\equiv& \frac{1}{2 (2\pi)^2} \left( \frac{\omega_a}{c} \right)^3 \frac{d^2}{\hbar \epsilon_0} \Lambda,
\eeqa
where the integration runs over the solid angle ${\rm d}\Omega$ covered by the pulse mode and the indices ${\bf k}_a$ respect the condition $|{\bf k}_a| = k_a$ coming from the delta distribution in \eq{\ref{eq Gamma_p}}. Note, that in the special case, where $u_{{\bf k}_a,\lambda} = {\rm e}^{i \,{\bf k}_a \cdot {\bf r}_a}$ and the integration is performed over the whole solid angle, the parameter lambda reaches its maximum value $\Lambda = 8\pi/3$ and one obtains the well known formula for the spontaneous decay of a single atom in a free space (see e.g. \cite[p. 530]{Tannoudji_2004})
\beq
	\Gamma = \frac{1}{3\pi} \left( \frac{\omega_a}{c} \right)^3 \frac{d^2}{\hbar \epsilon_0}.
\eeq
\noindent This is as well the maximum possible value (in principle) for the pulse mode decay $\Gamma_p = \Gamma \frac{\Lambda}{8\pi/3}$. Thus the single parameter $\Lambda \in [0, 8\pi/3]$ describes which part of space, weighted by the atomic dipole moment, is covered by the pulse.

In the following, we study the excitation probability $P_e$ of the atom excited by the photon wave-packet, which is given by the expectation value of atomic operator $\hat{\sigma}_{z}$
\begin{eqnarray}
	P_e(t)=\frac{1}{2}\Big(\langle \Psi_0|\hat{\sigma}_{z}(t)|\Psi_0\rangle+1\Big),
\end{eqnarray}
where the initial state of the system $|\Psi_0\rangle=|g\rangle |\Phi_{p}\rangle  |0_e\rangle$ is a product state of the atomic ground state, the pulse state, and the environment. We assume the environment to be initially in the vacuum state. In order to find the value of $\left< \hat{\sigma}_z(t) \right>$ at arbitrary time $t$, one has to solve the set of coupled differential equations for all involved time-dependent operators ( in our case $\hat{\sigma}_z$ and $\hat{\sigma}_\pm$, where the equation for $\hat{\sigma}_+$ is just a hermitian conjugate of \eq{\ref{c1}} ). The complete set of equations is obtained by first averaging the Eq. (\ref{z1}) over the initial state $\ket{\Psi_0}$. Next, the initial state dependent average of operators $\hat{\sigma}_\pm$ has to be found and can be obtained from Eqs. (\ref{c1}) and H.c. of (\ref{c1}). The complete set of equations can be schematically written as
\begin{eqnarray}
	\dot{\bf s}(t)&=&M\, {\bf s}(t)+{\bf b}.
	\label{eq set}
\end{eqnarray}
The form of the vectors ${\bf s}$, ${\bf b}$ and the matrix $M$ is initial state dependent and will be specified in Sec. \ref{s3}. Note that the Langevin-type noise operators \eq{\ref{eq zeta_z},\ref{eq zeta_-}} are determined directly in terms of the initial field operators of the environment \cite[p. 273]{Scully_1997}, so their average values will vanish as $< \hat{\zeta} >$=0, when considering the initial vacuum state of the environment.


\section{Special case: dipole pattern}
\label{s3}
With the help of the presented general model, one can study the dependence of excitation probabilities on both the spatial and temporal properties of the pulse. In the following, we assume that the light field in question matches the atomic dipole field pattern. Therefore, all the functions $g_{{\bf k},\lambda}$ will be defined with the dipole pattern. Then we only focus on the temporal and spectral effects of pulse. We first examine the interaction of the atom with a single photon Fock state pulse and next with the coherent state pulse.
\subsection{Single photon wave-packet}
Let's first consider the pulse mode to be a single photon wave-packet, which can be written as \cite[p. 208]{Scully_1997}
\beqa
	\ket{1_p} &=& \sum_{\lambda} \int d^3 {\bf k} g_{{\bf k},\lambda}^* ({\bf r}_a) f(\omega_k) \hat{a}^\dagger_{{\bf k},\lambda} \ket{0} \nonumber \\
						&\equiv& \hat{a}^\dagger_{C}\ket{0}						
	\label{eq single photon}
\eeqa
where the spectral distribution function $f(\omega_k)$ is the only degree of freedom left. The normalization of the single photon wave-packet \eq{\ref{eq single photon}} implies,
  \beqa
  \sum_{\lambda} \int d^3 {\bf k} {|g_{{\bf k},\lambda}({\bf r}_a)|}^2 {|f(\omega_k)|}^2 = 1.
  \eeqa
   The explicit form of \eq{\ref{eq set}} can now be found with the initial state $\ket{\Psi_0} = \ket{g} \ket{1_p} \ket{0_e}$: \\
\[
	{\bf s}(t)= \left(
	\begin{array}{ccc}
			\langle g, 1_p, 0_e|\,\hat{\sigma}_{z}(t)\,|g, 1_p, 0_e\rangle \\
			\langle g, 1_p, 0_e|\,\hat{\sigma}_{+}(t)\,|g, 0_p, 0_e\rangle \\
			\langle g, 0_p, 0_e|\,\hat{\sigma}_{-}(t)\,|g, 1_p, 0_e\rangle
	\end{array} \right)
\]
\begin{equation}
	M = \left(
	\begin{array}{ccc}
			-\Gamma & -2g(t) & -2g^*(t) \\
			0 & -\Gamma/2 & 0 \\
			0 & 0 & -\Gamma/2
	\end{array} \right),
\qquad
	{\bf b} = \left(
	\begin{array}{ccc}
			-\Gamma \\
			-g^*(t) \\
			-g(t)
	\end{array} \right),
\end{equation}
with initial condition
\[
	{\bf s}^T(t_0)= \left(
	\begin{array}{ccc}
			-1 & 0 & 0
	\end{array} \right).
\]

\noindent Once again with the help of Weisskopf-Wigner approximation, where we assume that the coupling $g_{{\bf k},\lambda}$ is constant for frequencies of interest centered around the atomic transition frequency $\omega_a$, one finds that the effective coupling strength $g(t)$ is the product of temporal envelope of the pulse and the decay rate into the pulse mode,
\begin{eqnarray}
	g(t)&=& \sqrt{\Gamma_p} \, \xi(t),
	\label{eq gt}
\end{eqnarray}
where
\begin{eqnarray}
	\xi(t)&=&\frac{\sqrt{\Gamma_p}}{2 \pi} \int d\omega_k f(\omega_k)e^{-i(\omega_k-\omega_a)t},
\end{eqnarray}
is up to a constant factor the Fourier transform of the spectral distribution function.
The excitation probability is then given by the first component of the vector ${\bf s}$
\begin{eqnarray}
	P_e(t)=\frac{1}{2}(s_1(t)+1).
\label{p1}
\end{eqnarray}

\subsection{Coherent state wave-packet}
Let us now have a look at the wave packet initially prepared in a (continuous field) coherent state. In analogy with the definition of continuous mode coherent state presented in \cite{Blow_1990}, we define the coherent state as
\beq
	\ket{\alpha_p} =  {\rm exp} \left[\alpha \hat{a}_{C}^\dagger -\alpha^* \hat{a}_{C} \right] \ket{0},
\eeq
where the wave-packet operator $\hat{a}_{C}^\dagger$ is defined in \eq{\ref{eq single photon}}. The mean photon number $N$ in the wave packet is given by,
\beq
N = \bra{\alpha_p} \hat{a}_{C}^\dagger \hat{a}_{C} \ket{\alpha_p} = {|\alpha|}^2.
\eeq
This yields the usual relation for coherent state
\beq
	\hat{a}_{C} \ket{\alpha_p} = \alpha \ket{\alpha_p}.
\eeq
Again, we get a set of similar differential equations with the same initial condition, but different variables
\[
	{\bf s}(t)= \left(
	\begin{array}{ccc}
		\langle g, \alpha_p, 0_e|\,\hat{\sigma}_{z}(t)\,|g, \alpha_p, 0_e\rangle \\
		\langle g, \alpha_p, 0_e|\,\hat{\sigma}_{+}(t)\,|g, \alpha_p, 0_e\rangle \\
		\langle g, \alpha_p, 0_e|\,\hat{\sigma}_{-}(t)\,|g, \alpha_p, 0_e\rangle
	\end{array} \right),
\]

\begin{equation}
	M = \left(
	\begin{array}{ccc}
			-\Gamma & -2g(t) & -2g^*(t) \\
			g^*(t) & -\Gamma/2 & 0 \\
			g(t) & 0 & -\Gamma/2
	\end{array} \right),
\qquad
	{\bf b} = \left(
	\begin{array}{ccc}
			-\Gamma \\
			0 \\
			0
	\end{array} \right).
\end{equation}


\section{Analysis of the temporal envelope}
\label{s4}
In this section, we assume that the incoming pulse not only matches the atomic dipole pattern but also occupies the whole solid angle, which implies that $\Gamma_p = \Gamma$. We then discuss the pulse temporal shape effect on the excitation probability.

\subsection{Pulse bandwidth effects}
\begin{figure}[h!]
\includegraphics[scale=0.3]{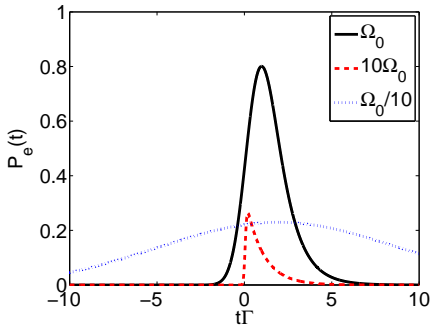}
\caption{(Color online) Excitation probability $P_e(t)$ as a function of time with the initial Gaussian pulse in single photon Fock state for different bandwidths. $\Omega_0=1.5\Gamma$, which turns out to be the optimized bandwidth (c.f. Fig.\ref{width2}).}
\label{width1}
\end{figure}

For a fixed pulse envelope, the excitation probability depends on ratio between the pulse bandwidth $\Omega$ and the decay rate $\Gamma$ of the atomic dipole. We take a single photon Fock state pulse with a Gaussian temporal shape as an example, and study the effects of different bandwidths on the excitation probability. The results are plotted in Fig.\ref{width1}.

As we can see from Fig.\ref{width1}, for single photon excitation with shorter pulses ($\Omega\gg\Gamma$), the bandwidth is too broad for resonant absorption, which reduces the effective coupling strength. For longer pulses ($\Omega\ll\Gamma$), the photon density is too low for efficient interactions  \cite{Domokos_2002}.

\begin{figure}[h!]
\includegraphics[scale=0.3]{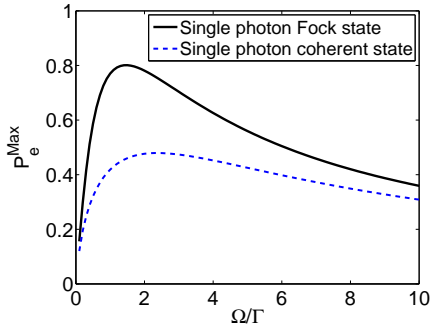}
\caption{(Color online) Dependence of maximum excitation probability $P^{\text{Max}}_e$ on the pulse bandwidth with Gaussian shape for single photon Fock state pulse and single photon coherent state pulse.}
\label{width2}
\end{figure}

In Fig.\ref{width2}, we show the dependence of the maximum achievable resonant excitation probability on the pulse bandwidth for the single photon Fock state pulse and single photon coherent state pulse, where the mean photon number equals 1. We find out that the optimum pulse bandwidth maximizing the absorption is $\Omega_0=1.5\Gamma$ for single photon Fock state pulse and $\Omega'_0=2.4\Gamma$ for single photon coherent state pulse.

\begin{figure}[h!]
\includegraphics[scale=0.3]{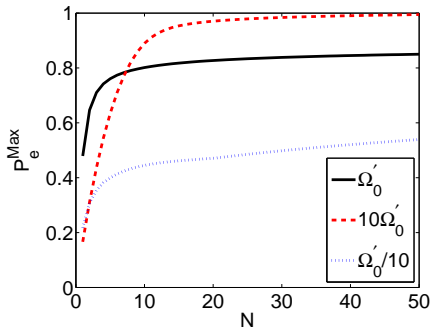}
\caption{(Color online) Maximum excitation probability $P^{\text{Max}}_e$ as a function of the mean photon number $N$ with the initial coherent state Gaussian pulse for different bandwidths. $\Omega'_0=2.4\Gamma$ is the optimized bandwidth for a Gaussian pulse (c.f. Fig.\ref{width2}).}
\label{width3}
\end{figure}

For coherent state pulses, we studied the maximal excitation probability as a function of the mean number of photons N for various choices of the bandwidth, shown in Fig.\ref{width3}. As expected, the maximal excitation probability varies with $N$. The saturation for large $N$ for all bandwidths is due to the fact that the effective coupling strength $g(t)$ decreases with the pulse length.  Alternatively, this can be understood as the photons arrive more distributed in time. Note that, for large $N$, it is better to choose short intense pulse with $\Omega\gg\Omega'_0\sim\Gamma$, which is used for population transfer.

\subsection{Pulse shape effects}
In general, the excitation probability depends on the specific shape of the input pulse. Here we studied the following six pulse shapes, see Table \ref{tab1}.
\begin{table}[h!]
\extrarowheight=3pt
\caption{Definition of pulse shapes}
\label{tab1}
\begin{tabular*}{8.5cm}{lc}
  \hline\hline
  Type of pulse & Wave function for pulse bandwidth  \\
  \hline
  \\[-6pt]
  Gaussian pulse & $\xi(t)={(\frac{\Omega^2}{2 \pi})}^{1/4} \exp\left(-\frac{\Omega^2}{4}\,t^2\right) $  \\
  \\[-6pt]
   Hyperbolic secant pulse & $\sqrt{\frac{\Omega}{2}}\, \text{sech}{(\Omega\,t)}$ \\
  \\[-6pt]
  Rectangular pulse & $\xi(t)=\left\{
\begin{array}{cc}
 \sqrt{\frac{\Omega}{2}},  & \text{for} \,0 \leq t \leq \frac{2}{\Omega} \\
 0,  & else
\end{array}
\right.$ \\
\\[-6pt]
 Symmetric exponential pulse & $\xi(t)=\sqrt{\Omega} \exp\left(-\Omega \,|t|\right)$ \\
  \\[-6pt]
  Decaying exponential pulse & $\xi(t)=\left\{
\begin{array}{cc}
 \sqrt{\Omega} \exp\left(-\frac{\Omega}{2}\,t\right), & \text{for} \, t > 0 \\
 0,  &  \text{for} \, t < 0
\end{array}
\right.$ \\
\\[-6pt]
  Rising exponential pulse & $\xi(t)=\left\{
\begin{array}{cc}
 \sqrt{\Omega} \exp\left(\frac{\Omega}{2}\,t\right), & \text{for} \,t < 0 \\
 0,  &  \text{for} \, t > 0
\end{array}
\right.$  \\
\\[-6pt]
  \hline\hline
  \end{tabular*}
\end{table}

\begin{figure}[h!]
\begin{minipage}{0.49\linewidth}
  \leftline{\includegraphics[width=4.34cm]{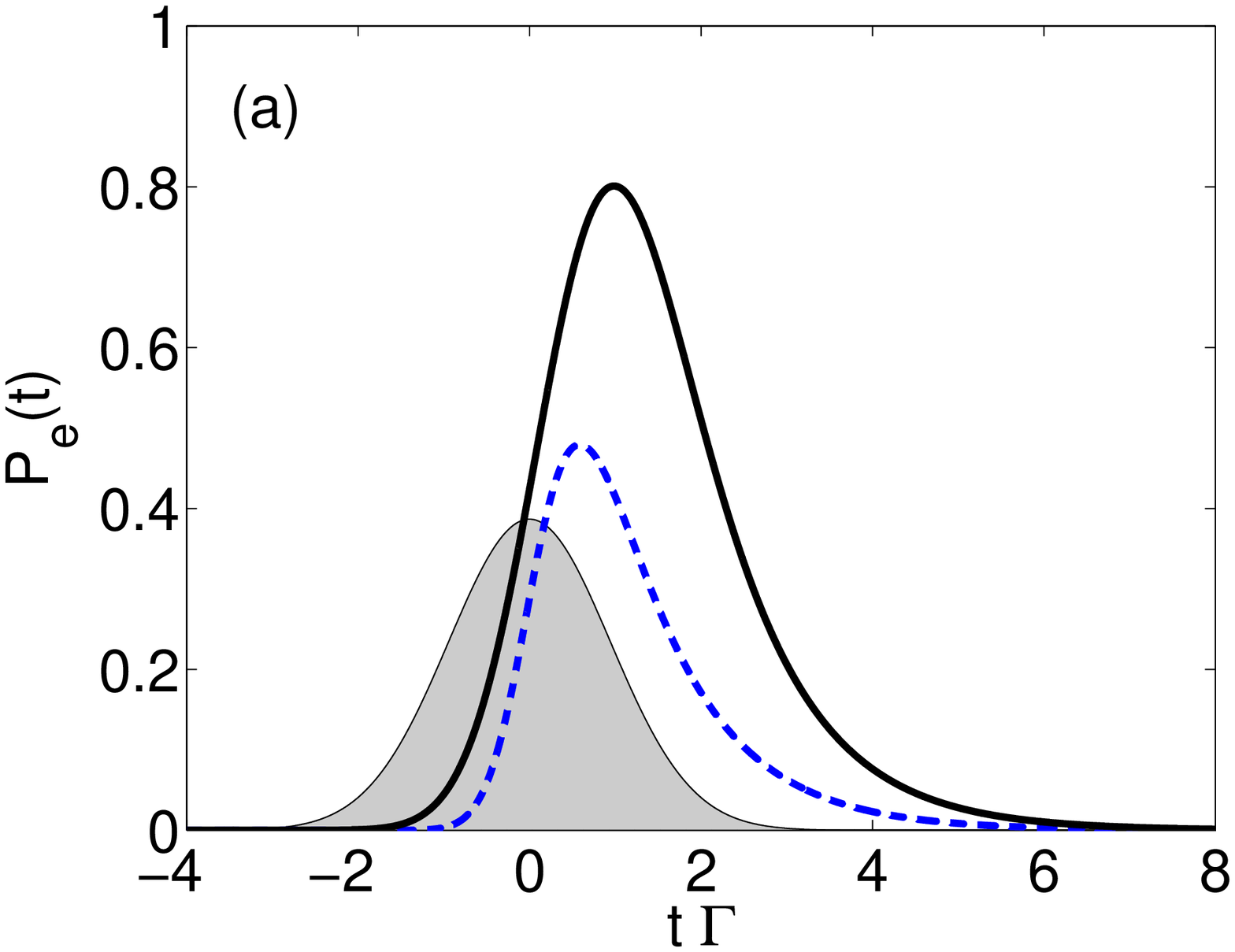}}
\vspace{0.0cm}
\end{minipage}
\hfill
\begin{minipage}{0.49\linewidth}
  \rightline{\includegraphics[width=4.34cm]{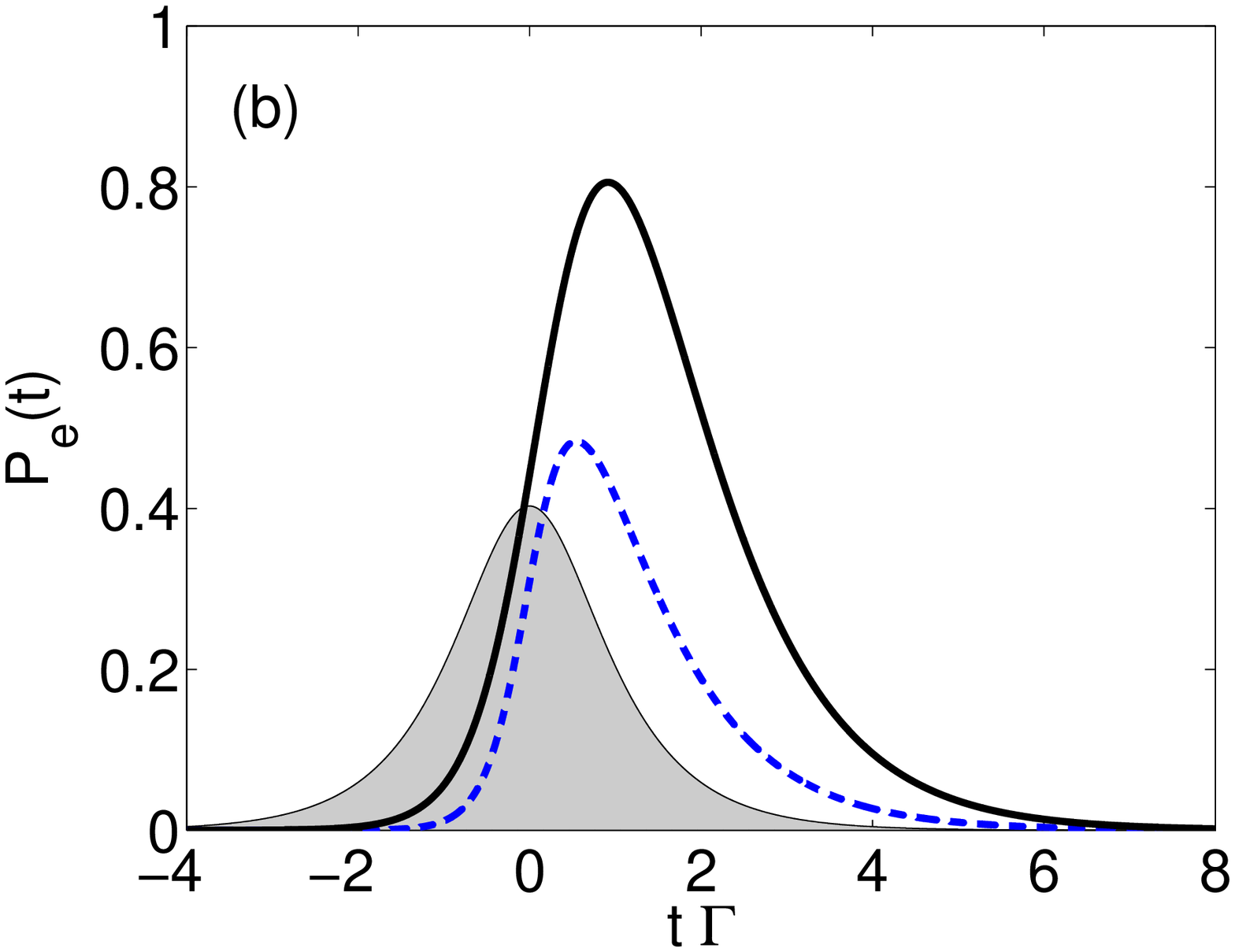}}
\vspace{0.0cm}
\end{minipage}
\begin{minipage}{0.49\linewidth}
  \leftline{\includegraphics[width=4.34cm]{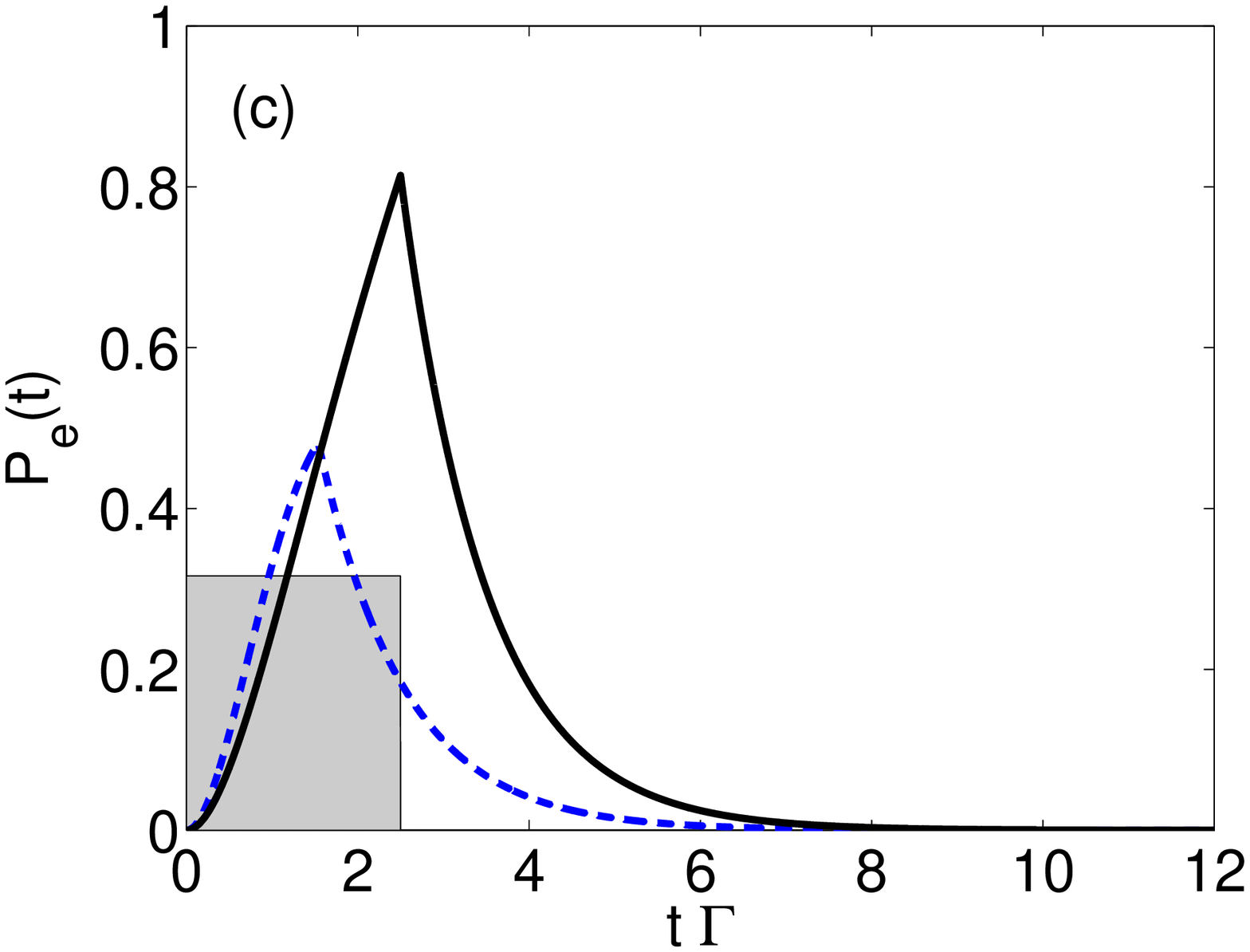}}
\vspace{0.0cm}
\end{minipage}
\hfill
\begin{minipage}{0.49\linewidth}
  \rightline{\includegraphics[width=4.34cm]{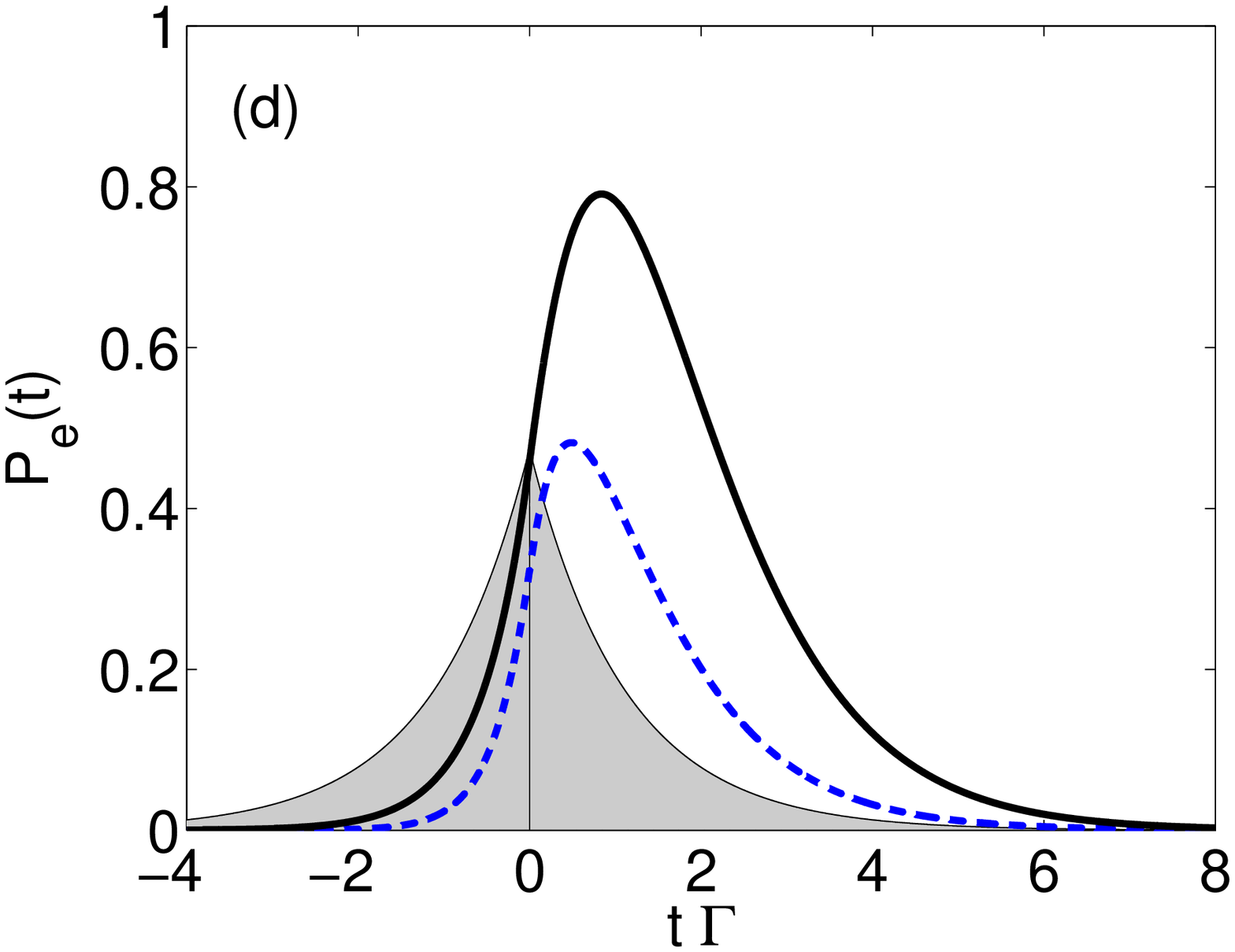}}
\vspace{0.0cm}
\end{minipage}
\begin{minipage}{0.49\linewidth}
  \leftline{\includegraphics[width=4.34cm]{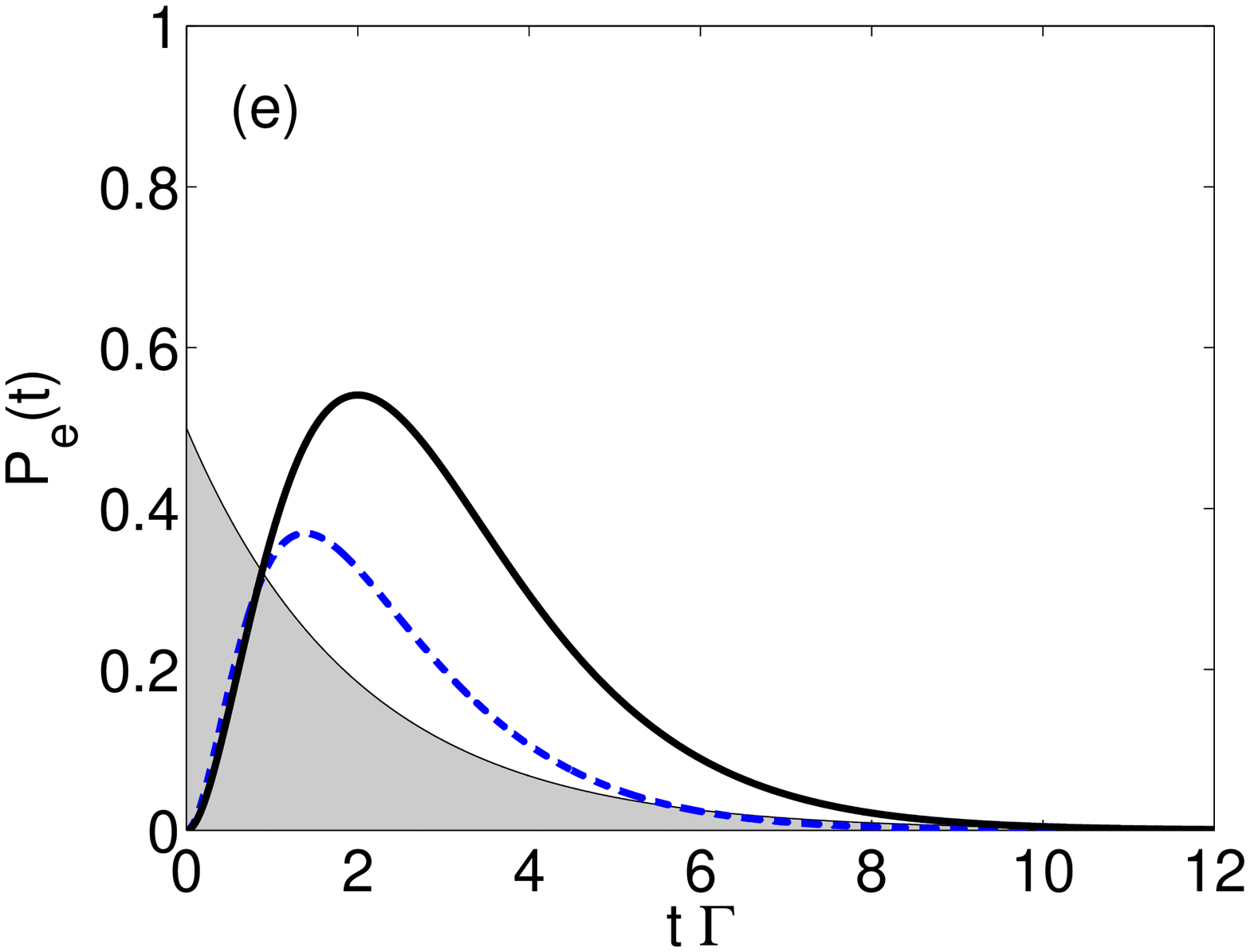}}
\vspace{0.0cm}
\end{minipage}
\hfill
\begin{minipage}{0.49\linewidth}
  \rightline{\includegraphics[width=4.34cm]{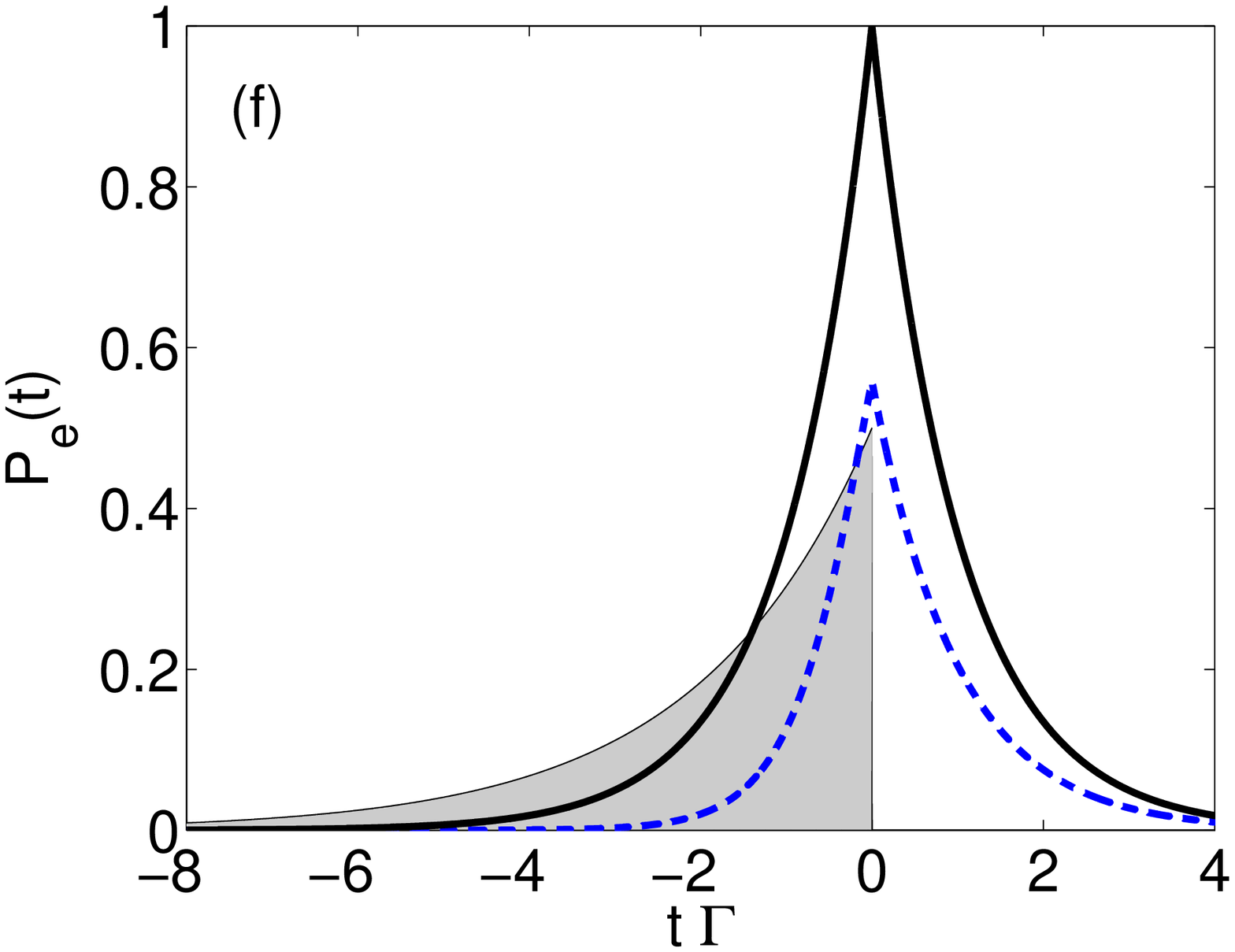}}
\vspace{0.0cm}
\end{minipage}
\caption{(Color online) Excitation probability $P_e(t)$ as a function of time for the coupling strength $g(t)$ given in Eq.(\ref{eq gt}) with $\Gamma_p = \Gamma$ ($\Lambda=8\pi/3$). The single photon Fock state pulse with optimal bandwidth is shown in grey; the corresponding excitation probability is given by the solid black line. The dashed blue line represents the excitation probability for a single photon coherent state pulse of a similar shape but different (optimized) bandwidth. (a) Gaussian pulse, (b) Hyperbolic secant pulse, (c) Rectangular pulse, (d) Symmetric exponential pulse, (e) Decaying exponential pulse and (f) Rising exponential pulse.}
\label{fig1}
\end{figure}

For single photon Fock state, the excitation probability has a peak value of about $0.8$ with optimum bandwidth for the first four pulse shapes, shown in Fig.\ref{fig1} (a)-(d), indicating that the photon absorption is less sensitive to pulse shape effects such as discontinuities. For the decaying exponential pulse, the maximum excitation probability is only $0.54$, see Fig.\ref{fig1} (e). A particularly interesting case may be that of the rising exponential single photon Fock state pulse, shown in Fig.\ref{fig1} (f), for which the corresponding maximal excitation probability is $0.995$ with a optimal bandwidth of $\Omega_0=1\Gamma$. This agrees well with the prediction that for the aim of unit absorption probability, the incident photon must possess the time reversed properties of the spontaneously emitted photon. Since the spontaneous decay is exponential, the temporal envelope of the pulse has to be rising exponential \cite{Stobinska_2009,Heugel_2010}.

On the other hand, for an initial single photon coherent state pulse with optimum bandwidth, the maximum excitation probability is much lower, around $0.48$ for the first four pulse shapes and $0.4$ and $0.56$ for the decaying and rising exponential pulse, respectively. Apparently the excitation is more efficient if exactly one photon is present instead of a distribution with mean one. This emphasis the importance of generating single photon source rather than using attenuated laser pulse in applications where a high absorption is desired.

For the explicit values of optimum bandwidth needed to achieve maximum excitation probability, see Table. \ref{t2}

\begin{table}[h!]
\caption{Optimum bandwidth and maximum excitation probability, the results $(*)$ were also obtained in Ref.\cite{Stobinska_2009} with a different method.}
\label{t2}
\begin{tabular*}{8.75cm}{l|c|c|c}
\hline\hline
Tape of pulse& State& Optimum $\Omega/\Gamma$& Maximum $P_e(t)$\\
\hline
\multirow{2}*{Gaussian pulse} &$|\alpha\rangle$ &$2.4$  &$0.48$\\
\cline{2-4}
&$|1\rangle$ &$1.5$  &$0.80$ *\\
\hline
\multirow{2}*{Hyperbolic secant pulse} & $|\alpha\rangle$ &$2.0$  &$0.48$\\
\cline{2-4}
&$|1\rangle$ &$1.3$ &$0.80$\\
\hline
\multirow{2}*{Rectangular pulse} & $|\alpha\rangle$ &$1.3$ &$0.48$\\
\cline{2-4}
&$|1\rangle$ &$0.8$ &$0.81$\\
\hline
\multirow{2}*{Symmetric exponential pulse} & $|\alpha\rangle$ &$1.4$  &$0.48$\\
\cline{2-4}
& $|1\rangle$ &$0.9$  &$0.79$\\
\hline
\multirow{2}*{Decaying exponential pulse} & $|\alpha\rangle$ &$1.4$ &$0.37$\\
\cline{2-4}
& $|1\rangle$ &$1.0$ &$0.54$ *\\
\hline
\multirow{2}*{Rising exponential pulse} & $|\alpha\rangle$ &$1.9$  &$0.56$\\
\cline{2-4}
&$|1\rangle$ &$1.0$ &$0.995$ *\\
\hline\hline
\end{tabular*}
\end{table}

\subsection{Damped Rabi oscillation}

\begin{figure}[h!]
\includegraphics[scale=0.3]{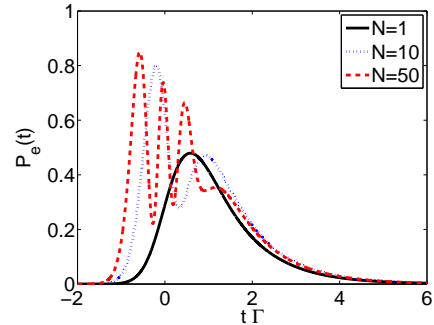}
\caption{(Color online) Excitation probability $P_e(t)$ as a function of time for initial coherent state Gaussian pulse with optimal bandwidth $\Omega'_0=2.4\Gamma$ for different mean photon numbers $N$.}
\label{damp}
\end{figure}

In Fig.\ref{damp}, the probability of exciting the atom for an initial coherent state Gaussian pulse is evaluated for various mean photon numbers $N=(1,10,50)$. For large mean photon number, damped Rabi oscillations are observed. In the limit of very large mean photon number, one would recover the textbook predictions for classical light pulses"  \cite[p. 151]{Scully_1997}.

\section{Discussion of realistic focusing}
\label{s5}

Finally we present a brief review of ongoing experiments in order to consider the excitation probability in realistic tight focusing configurations.

In the case of a parabolic mirror with a half opening angle of $134^\circ$ as it is used in the experiment described in Refs.\cite{Lindlein_2007,Sondermann_2007}, the corresponding weighted solid angle reaches $\Lambda=0.94\times8\pi/3$, and thus one may achieve a maximal excitation probability of $0.94$ with rising exponential shape for a single photon Fock state pulse, $0.54$ for a single photon coherent state pulse, $0.75$ for a Gaussian single photon Fock state pulse and $0.46$ for a single photon coherent state pulse.

In Ref.\cite{Tey_2008,Tey_2009}, a high aperture lens with $NA=0.55$ and $f=4.5$ mm is used to focus down a Gaussian beam. The weighted solid angle depends on the focusing strength $u:=w_L/f$, where $w_L$ is the beam waist. A maximum overlap of $\Lambda=0.364\times8\pi/3$ is expected at focusing strength $u=2.239$. With a rising exponential shape, we predict a maximal excitation probability of 0.36 for a single photon Fock state pulse and 0.27 for a single photon coherent state pulse. For a Gaussian shape, we predict a maximal excitation probability of 0.29 for a single photon Fock state pulse and 0.23 for a single photon coherent state pulse.

\section{Conclusion}
\label{s6}
In conclusion, with the help of time dependent Heisenberg-Langevin equations, we studied the interaction between a single two-level atom and a propagating pulse at the quantum level. We have presented a general approach and a scalar model to treat the atom-pulse interaction. For strong focusing configurations we account for the overlap of the incoming pulse mode with the respective atomic dipole pattern.

The effect of temporal-spectral features of the single photon Fock state pulse and coherent state pulse on excitation probability of the atom has been investigated. With Gaussian, hyperbolic secant, rectangular and symmetric exponential shape pulses, the achievable maximum excitation probability is around $0.8$ for single photon Fock state pulse and $0.48$ for single photon coherent state pulse. More importantly, with a rising exponential shape, the maximum value is nearly $1$ for a single photon Fock state pulse and $0.55$ for a single photon coherent state pulse, which is in agreement with the time reversal argument. As an example, the effect of bandwidths and mean photon numbers of a Gaussian pulse is analyzed. We also survey some current arrangements designed to couple photons and atoms in order to assess their potential for high excitation probabilities. The presented model provides a suitable foundation to further study the pulse shape effects of quantized continuous light fields in both scalar and full three-dimensional treatment.

\section{Acknowledgements}
We would like to thank Colin Teo, Syed Abdullah Aljunid, Gleb Maslennikov and Christian Kurtsiefer for useful discussions. This work was supported by the National Research Foundation and the Ministry of Education, Singapore.

\bibliographystyle{prsty}
\bibliography{yiminBib0}

\end{document}